\documentclass[aps, twocolumn, superscriptaddress, floatfix]{revtex4-2} 

\usepackage{graphicx}
\usepackage[utf8]{inputenc}
\usepackage{amssymb, amsmath, commath, braket, physics, soul}
\usepackage{hyperref}
\hypersetup{
	colorlinks=true,
	linkcolor=red,
	filecolor=blue,
	urlcolor=magenta,
}
\usepackage{placeins, standalone}

\newcommand\Hlr{\ensuremath{\hat{H}_\mathrm{LR}}}

\newcommand\mean[1]{\ensuremath{\langle #1 \rangle}}

\newcommand\tTh{\ensuremath{t_\mathrm{dip}}}
\newcommand\tH{\ensuremath{t_\mathrm{H}}}

\begin{document}
	
\title{Proposal for many-body quantum chaos detection}
\author{Adway Kumar Das}
\affiliation{Department of Physics, University of Connecticut, Storrs, Connecticut 06269, USA}
\affiliation{Department of Physical Sciences, Indian Institute of Science Education and Research Kolkata, Mohanpur 741246, India}
\author{Cameron Cianci}
\affiliation{Department of Physics, University of Connecticut, Storrs, Connecticut 06269, USA}
\affiliation{Mirion Technologies (Canberra) Inc., 800 Research Parkway, Meriden, Connecticut 06450, USA}
\author{Delmar G. A. Cabral}
\affiliation{Department of Chemistry, Yale University, P.O. Box 208107, New Haven, Connecticut 06520-8107, USA}
\author{David A. Zarate-Herrada}
\affiliation{Institut of Physics, Benem{\'e}rita Universidad Aut{\'o}noma de Puebla, Puebla, 72570, Mexico}
\author{Patrick Pinney}
\affiliation{Department of Physics, University of Connecticut, Storrs, Connecticut 06269, USA}
\author{Sa\'ul Pilatowsky-Cameo}
\affiliation{Center for Theoretical Physics, Massachusetts Institute of Technology, Cambridge, Massachusetts 02139, USA}
\author{Apollonas S. Matsoukas-Roubeas}
\affiliation{Department of Physics and Materials Science, University of Luxembourg, L-1511 Luxembourg, G. D. Luxembourg}
\author{Victor S. Batista}
\affiliation{Department of Chemistry, Yale University, P.O. Box 208107, New Haven, Connecticut 06520-8107, USA}
\author{Adolfo del Campo}
\affiliation{Department of Physics and Materials Science, University of Luxembourg, L-1511 Luxembourg, G. D. Luxembourg}
\affiliation{Donostia International Physics Center,  E-20018 San Sebasti\'an, Spain}
\author{E. Jonathan Torres-Herrera}
\affiliation{Institut of Physics, Benem{\'e}rita Universidad Aut{\'o}noma de Puebla, Puebla, 72570, Mexico}
\author{Lea F. Santos}
\affiliation{Department of Physics, University of Connecticut, Storrs, Connecticut 06269, USA}

\begin{abstract}
In this work, the term ``quantum chaos'' refers to spectral correlations similar to those found in the random matrix theory. Quantum chaos can be diagnosed through the analysis of level statistics using e.g.~the spectral form factor, which detects both short- and long-range level correlations. The spectral form factor corresponds to the Fourier transform of the two-point spectral correlation function and exhibits a typical slope-dip-ramp-plateau structure (aka correlation hole) when the system is chaotic. We discuss how this structure could be detected through the quench dynamics of two physical quantities accessible to experimental many-body quantum systems: the survival probability and the spin autocorrelation function. The survival probability is equivalent to the spectral form factor with an additional filter. When the system is small, the dip of the correlation hole reaches sufficiently large values at times which are short enough to be detected with current experimental platforms. As the system is pushed away from chaos, the correlation hole disappears, signaling integrability or localization. We also provide a relatively shallow circuit with which the correlation hole could be detected with  commercially available quantum computers.  
\end{abstract}

\maketitle
\section{Introduction}

The main mechanism for the onset of quantum chaos in many-body quantum systems is the interaction between particles. Similar to what one finds in random matrix theory, realistic many-body quantum systems in the chaotic regime are characterized by correlated energy levels~\cite{Brody1981,ZelevinskyRep1996} and middle-spectrum eigenstates that approach random vectors by filling the energy shell~\cite{Borgonovi2016}.
When these systems are taken far from equilibrium, quantum chaos  underlies the spread and scrambling of quantum information, hindering the reconstruction of the initial state through local measurements. 
Such redistribution of quantum information is intertwined with the thermalization of subsystems~\cite{Borgonovi2016,Alessio2016} and the difficulty in reaching a localized phase~\cite{Luitz2017,Suntajs2020,Abanin2021}.
Understanding and quantifying many-body quantum chaos is thus essential for describing and controlling many-body quantum dynamics and for the development of quantum technologies.  In this work, we discuss how spectral correlations indicative of quantum chaos can be experimentally detected via the dynamics of many-body quantum systems. 

The spectral form factor provides direct access to  short- and long-range correlations among the eigenvalues of a system Hamiltonian. Mathematically, it is defined as the Fourier transform of the two-point function of the energy spectrum~\cite{MehtaBook}. In chaotic systems, it presents a slope-dip-ramp-plateau structure analogous to the one found in random matrix theory, therefore signaling a rigid spectrum. This structure can persist even in the presence of environmental noise~\cite{Tameshtit1992,Xu2021,Cornelius2022,Cao2022,Matsoukas2023,Matsoukas2023pra,Matsoukas2023quantum, VallejoFabila2024}. The ramp only appears when the levels are correlated and the plateau represents the saturation value of the spectral form factor.  The ramp reaches the plateau at the Heisenberg time, which is inversely proportional to the mean level spacing and thus depends on the dimension of the Hilbert space.  

The analysis of level statistics through the spectral form factor is an excellent way to detect many-body quantum chaos in experiments with access to the spectrum, as in nuclear physics. Level statistics is a less efficient diagnostic tool of chaos in experiments with cold atoms~\cite{Kinoshita2006,  Kaufman2016,Bordia2016,Hild2014,Schreiber2015,Rispoli2019}, ion traps  \cite{Jurcevic2014, Richerme2014,Smith2015,Kranzl2023} and available quantum computers~\cite{IBM,Smith2019}, where the spectrum is not easily accessible and the focus is instead on many-body quantum dynamics. To detect the slope-dip-ramp-structure through dynamics, recent works have proposed to monitor the fidelity of thermofield double states evolved under the Sachdev-Ye-Kitaev model~\cite{Xu2021} and to use measurement protocols on evolved random product states \cite{Joshi2022, Vasilyev2020} that are amenable to spin models realizable in platforms of Rydberg atoms~\cite{Browaeys2020},  superconducting qubits~\cite{Kjaergaard2020} and  stroboscopically-driven cold atoms in optical lattices~\cite{Dag2023}.

Our approach to detect many-body quantum chaos is suitable to different platforms which study dynamics, including digital and analog quantum computers. We propose the use of two dynamical quantities that can be experimentally measured and that, like the spectral form factor, exhibit the characteristic slope-dip-ramp-plateau structure when the system is chaotic. They are the survival probability and the spin autocorrelation function. 

The survival probability is defined as the squared absolute value of the overlap between the initial state and its evolved counterpart. It is equivalent to the spectral form factor with an additional filter associated with an initial state. We investigate the survival probability of experimentally accessible initial states.

The idea of detecting quantum chaos through the survival probability was first proposed in~\cite{Leviandier1986}, where the slope-dip-ramp-plateau structure was originally known as the ``correlation hole''~\cite{Leviandier1986,Pique1987,Guhr1990,Hartmann1991,Alhassid1992,Lombardi1993,Michaille1999,Leyvraz2013,Torres2017Philo,Torres2017,Torres2018,Schiulaz2019,Lerma2019,Santos2020,Xu2021,Lezama2021,Cao2022,Das2022a,Das2022b,Das2023,shir2023range}. 
It was later shown that the correlation hole emerges also in the spin autocorrelation function~\cite{Torres2018,Schiulaz2019}, which, contrary to the survival probability, is a local quantity in real space. As the parameters of the system are changed and it moves from the chaotic regime toward an integrable point or a localized phase, the correlation hole gradually disappears~\cite{Torres2017,Torres2017Philo,Torres2018,Schiulaz2019}.
 In this process, the correlation hole gets shallower and the ramp is postponed to longer times, as first described  in~\cite{Schiulaz2019} and then  in~\cite{Suntajs2020}.

In the present work, we study the evolution of the survival probability and the spin autocorrelation function in two different many-body spin-1/2 models that can be realized in current experiments with cold atoms, ion traps, nuclear magnetic resonance (NMR) platforms~\cite{Wei2018,Peng2023}, and in quantum computers. They are the one-dimensional (1D) disordered spin-1/2 Heisenberg model and the 1D disordered long-range Ising model in a transverse field.  When the disorder strength is comparable to the interaction strength, these systems are chaotic and present a correlation hole. When the disorder strength is large, the spectrum of finite systems ceases to be rigid and the correlation hole disappears. In this case, the dynamics present only oscillations.

The main challenges of our proposal lie on the minimum value of the correlation hole and the timescale for its appearance. The lowest value (timescale) decreases (increases) with the dimension of the Hilbert space, which in turn grows exponentially with the size of our many-body quantum systems.
Nevertheless, we show that the correlation hole can emerge even when the systems have only 6 or 8 sites. For such small chains, the dip happens at  sufficiently large values and the Heisenberg time is sufficiently short for the potential detection of many-body quantum chaos with current experimental capabilities. 

We also provide a quantum circuit and compare its results for the survival probability with those for the numerical simulation of the Heisenberg model with 6 sites. Due to the small chain size and the conservation of excitations in this model, the quantum circuit is relatively shallow, allowing for the implementation in current commercial quantum computers. We run our circuit in a fake provider that mimics the behavior of a real IBM quantum system with noise. The results indicate the presence of a correlation hole. 

In addition to the many-body spin models, we present results for a disordered spin-1/2 chain with a single excitation and nearest-neighbor coupling. This system is analogous to the one-particle Anderson model, thus being localized in the thermodynamic limit for an infinitesimal disorder strength. However, when the chain is finite, it can present level correlations that get manifested in the dynamics and could be experimentally detected.

We append that after the online appearance of our work, an experimental paper~\cite{Dong2024} confirmed the viability of our proposal. Using a superconducting quantum processor, 
the group was able to detect the correlation hole for a disordered spin-1/2 model and a Floquet system with 5 spins.

The paper is organized as follows. 
In Sec.~\ref{secII},  we review the definition and properties of the spectral form factor,  survival probability, and spin autocorrelation function.
In Sec.~\ref{secIII}, we analyze the dynamics of the survival probability and the spin autocorrelation function for the 1D disordered isotropic Heisenberg spin-1/2 model with nearest-neighbor couplings and the 1D disordered long-range Ising model in a transverse field. In Sec.~\ref{Sec:circuit}, we compare the exact diagonalization result for the survival probability evolving under the Heisenberg model with that obtained with our quantum circuit run in an IBM fake provider. 
In Sec.~\ref{secIV}, the analysis is extended to the  1D spin-1/2 model with a single excitation.
Conclusions are presented in Sec.~\ref{secV}.

\section{Dynamical indicators of many-body quantum chaos} \label{secII}

The two-point spectral form factor captures both short- and long-range correlations in the energy spectrum, thus providing a complete diagnostic of quantum chaos. This quantity has also been used to question the existence of a many-body localized phase~\cite{Suntajs2020} and in recent studies of scale-invariant critical dynamics~\cite{Hopjan2023}.  The two-point spectral form factor is defined as~\cite{MehtaBook},
\begin{align}
	\label{eq_SFF}
	\mathrm{SFF}(t) = \dfrac{1}{D^2} \left\langle \sum_{m, n}e^{i(E_m - E_n)t} \right\rangle,
\end{align}
where $\hbar =1$, $D$ is the dimension of the Hilbert space, $E_n$ represents the eigenvalues of the system, and  $\langle\cdot \rangle$ indicates an ensemble average. For large random matrices from the Gaussian orthogonal ensemble (GOE), we have~\cite{Das2025ArxivBRM,Torres2018,Schiulaz2019} 
\begin{align}
	\label{eq_SFF_GOE}
	\mathrm{SFF}(t) \simeq \frac{\mathcal{J}_1^2(2 \Gamma t)}{(\Gamma t)^2}-\frac{1}{D} b_2\left(\frac{2\Gamma t}{\pi D}\right) + \frac{1}{D},
\end{align}
where $\mathcal{J}_1 (t)$ is the Bessel function of the first kind, $\Gamma$ is the width of the semicircular density of states, and $b_2 (t)$ is the two-level form factor~\cite{MehtaBook},
\begin{align}
	\label{eq_b2}
	b_2(t) &= \begin{cases}
		1 - 2t + t\log\del{1 + 2t}, & t\leq 1,\\
		t\log \left( \dfrac{2t + 1}{2t - 1} \right) - 1, & t > 1 .
	\end{cases}
\end{align}

\begin{figure}[t]
    \centering
    \includegraphics[width=\columnwidth]{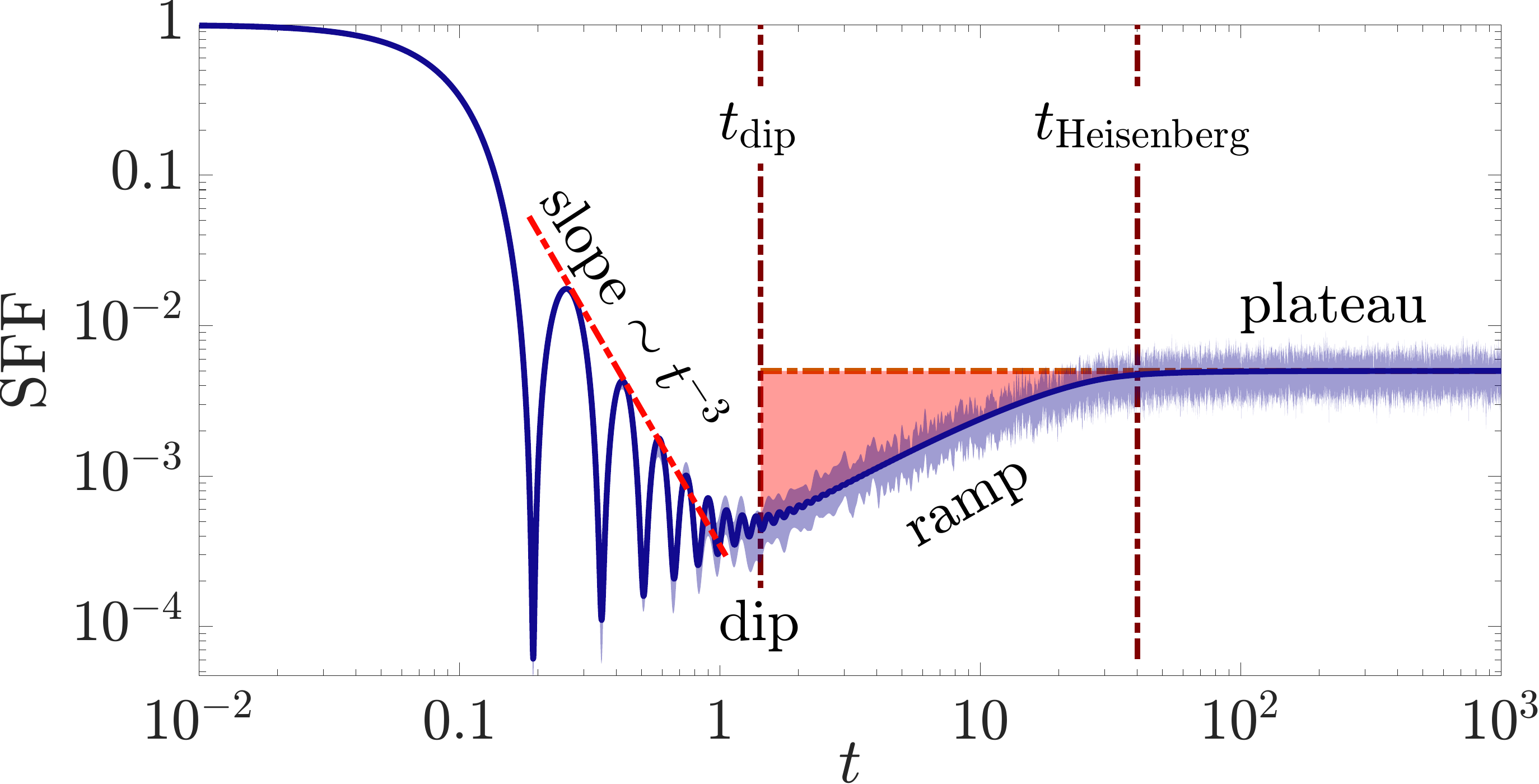}
    \caption{Slope-dip-ramp-plateau structure  (correlation hole) of the spectral form factor obtained with an ensemble of $100$ GOE random matrices of dimension $D=200$. The elements of the matrices have $\langle H_{ij}\rangle=0$, $\langle H_{ij}^2\rangle=1/2$ for $i\neq j$, and $\langle H_{ii}^2\rangle=1$, so that $\Gamma \sim \sqrt{D/2}$. The dark solid line is the analytical expression in Eq.~(\ref{eq_SFF_GOE}) and the shaded curve is the average over the ensemble. Vertical dashed lines mark the beginning and the end of the ramp. }
    \label{fig:GOE}
\end{figure}

In Fig.~\ref{fig:GOE}, we show the spectral form factor averaged over an ensemble of GOE random matrices.
The first term in Eq.~(\ref{eq_SFF_GOE}) represents the slope in the slope-dip-ramp-plateau structure. The slope exhibits oscillations characteristic of the Bessel function, whose envelope decays as a powerlaw $\propto t^{-3}$, as seen in Fig.~\ref{fig:GOE}. 

The dip in Fig.~\ref{fig:GOE} happens at $\tTh = 3^{\frac{1}{4}}$ and corresponds to the minimum value of the spectral form factor, $\mathrm{SFF}(\tTh) \approx (4/\pi) \sqrt{2/D^3}$ (see \cite{Das2025ArxivBRM,Schiulaz2019}). The dip is followed by the ramp, which is the region  below the saturation value (plateau) at $1/D$.  The beginning of the ramp and its end at the Heisenberg time, $t_{\text H}$, are marked with vertical lines in Fig.~\ref{fig:GOE}. The interval where $\mathrm{SFF}(t)$ is below the plateau corresponds to the correlation hole.

The ramp is described by the two-level form factor in Eq.~(\ref{eq_b2}) and it emerges due to spectral correlations. Unless averages are performed, the dip-ramp structure is hidden by fluctuations, because the spectral form factor is non-self-averaging~\cite{Prange1997,VallejoFabila2024,Schiulaz2020,Torres2020a,Torres2020b,Matsoukas2023pra}.

It is usual to add a filter to the spectral form factor, a common choice being the Boltzmann factors~\cite{Campo2018},
\begin{align}
	\label{eq:Boltzmann}
	f(E_n)= \frac{e^{-\beta E_n}}{\sum_m e^{-\beta E_m}},
\end{align}
 where $\beta$ is the inverse temperature. The case in Eq.~(\ref{eq_SFF}) is recovered for infinite temperature, $\beta=0$. 
 
\subsection{Survival probability}

The spectral form factor is closely related to the survival probability, which is defined as
\begin{equation}
	\label{eq_SP_def}
		S_P(t) = \abs{\braket{\Psi(0)}{\Psi(t)}}^2   =  \sum_{m,n} |c_m|^2 |c_n|^2 e^{i(E_m - E_n)t}  ,
\end{equation}
where $c_n = \braket{E_n}{\Psi(0)}$ is the $n$th component of the initial state $|\Psi(0) \rangle$ written in the energy eigenbasis $\{ \ket{E_n} \}$ of the Hamiltonian $\hat{H}$ that describes the system. If one equates the components $|c_n|^2$ with the  Boltzmann factors $f(E_n)$ in Eq.~(\ref{eq:Boltzmann}), then the spectral form factor can be interpreted as the survival probability of an initial Gibbs state~\cite{Campo2018}. However, this is not an easy state to prepare experimentally, so we focus instead on experimentally accessible initial states evolving under physical many-body quantum systems, as specified in Sec.~\ref{secIII}. 

We study quench dynamics, where the initial state is an eigenstate of the unperturbed Hamiltonian $\hat{H}_0$ and it evolves according to the total Hamiltonian,
\begin{equation}
    \hat{H} = \hat{H}_0 + \lambda\hat{V},
\end{equation}
where $\hat{V}$ is the perturbation and $\lambda$ is the strength of the perturbation. In many-body quantum systems with two-body couplings, as considered in this work, the density of states of $\hat{H}$ is Gaussian~\cite{Brody1981}. 

When the system is perturbed far from equilibrium $(\lambda \sim 1)$, the energy distribution of the initial state, which is often referred to as local density of states (LDOS),
\begin{align}
	\label{eq_LDOS}
	\rho_0(E) = \sum_{n = 1}^{D} |c_n|^2\delta (E - E_n),
\end{align}
is also Gaussian~\cite{Flambaum2000,Torres2014NJP}. The width $\Gamma$ of the LDOS is obtained as~\cite{Torres2016Entropy}
\begin{align}
	\label{eq_LDOS_width}
	\Gamma^2 &= \langle \Psi(0)|\hat{H}^2| \Psi(0) \rangle - \langle \Psi(0)|\hat{H}| \Psi(0) \rangle^2 \\
 &=
 \sum_{n \neq n_0}^D \abs{\bra{\varepsilon_n} \hat{H} \ket{\Psi(0)}}^2, \nonumber
\end{align}
where $|\varepsilon_n \rangle$ are the eigenstates of $\hat{H}_0$  and $n_0$ corresponds to the index of the initial state, $\ket{\Psi(0)}=\ket{\varepsilon_{n_0}}$. Notice that the calculation of $\Gamma$ only requires knowledge of the off-diagonal elements, $\bra{\varepsilon_n} \hat{H} \ket{\varepsilon_{n_0}}$, of the total Hamiltonian $\hat{H}$ written in the basis of eigenstates of $\hat{H}_0$.

The survival probability in Eq.~\eqref{eq_SP_def} can be equivalently expressed in terms of the Fourier transform of the LDOS,
\begin{align}
	\label{eq_SP_LDOS}
	S_P(t) &= \abs{\int_{E_\text{min}}^{E_\text{max}} \rho(E) e^{-iEt}dE}^2,
\end{align}
where $E_\text{min}$ and $E_\text{max}$ are the lower and upper energy bounds of the LDOS. In terms of Eq.~(\ref{eq_SP_LDOS}), it becomes clear that the initial decay of the survival probability is Gaussian, $e^{-\Gamma^2 t^2}$, followed by a power-law decay that presents oscillations~\cite{Tavora2016,Tavora2017,Campo2018a}. This is the slope part of the slope-dip-ramp-plateau structure. 

Since $S_P (t)$ is non-self-averaging~\cite{Schiulaz2020,Torres2020a},  we work with the averaged survival probability, $\langle S_P (t)\rangle$, to capture the subsequent features of the slope-dip-ramp-plateau structure, namely the dip and the ramp. In physical chaotic many-body quantum systems with time-reversal symmetry, $\langle S_P (t)\rangle$ reaches a minimum value at a time $\tTh \propto D^{2/3}/\Gamma$ \cite{Schiulaz2019}, after which the ramp emerges. The ramp is closely described by the $b_2(t)$ function in Eq.~(\ref{eq_b2}), and it persists up to the Heisenberg time, $t_{H} \propto D/\Gamma$ \cite{Schiulaz2019}. This is the largest timescale of the system.
Beyond $t_{H}$, the averaged survival probability exhibits small fluctuations around its infinite-time average,  
\begin{equation}
 \overline{ S_P } = \left\langle\sum_{n = 1}^{D} |c_n|^4 \right\rangle  .   
\end{equation} 

In chaotic many-body systems, $\tTh\ll t_{\text H}$, while for systems that approach a localized phase, $\tTh/t_{\text H} \rightarrow 1$. The use of this ratio to describe the delocalized-localized transition was discussed for disordered many-body quantum systems in~\cite{Schiulaz2019} and then in~\cite{Suntajs2020}.

\subsection{Spin autocorrelation function} 

The spin autocorrelation function is another observable that detects the correlation hole~\cite{Torres2018,Schiulaz2019}  and it is also non-self-averaging at long times~\cite{Schiulaz2020,Torres2020a,Torres2020b}. This quantity is defined as
\begin{align}
	\label{eq_Imb_def}
	I_z(t) &= \frac{1}{L} \sum_{k = 1}^{L} \bra{\Psi(0)} \hat{\sigma}_k^z e^{i \hat{H} t} \hat{\sigma}_k^ze^{-i \hat{H} t} \ket{\Psi(0)},
\end{align}
where $L$ is the number of sites of the spin-1/2 chains that we consider and $\hat{\sigma}_k^z$ is the Pauli $z$-operator acting on site $k$. We denote the asymptotic value of the average $\langle I_z(t) \rangle$ as $\overline{I_z}$. 

Like the survival probability, the spin autocorrelation function is nonlocal in time,  but contrary to the survival probability, it is local in real space. Another difference between the two quantities is that the correlation hole fades away for the spin autocorrelation function as the system size increases~\cite{Lezama2021}.  Upon choosing a N\'eel state as the initial state, $I_z(t)$ is comparable to the density imbalance experimentally probed in cold atoms \cite{Schreiber2015, Bordia2016}. In NMR platforms, the spin autocorrelation function can be directly measured for initially mixed states~\cite{Peng2023}.

For both quantities, $\langle S_P(t) \rangle$ and $\langle I_z(t) \rangle$, the dynamics need to resolve the discreteness of the spectrum for the emergence of the correlation hole. This explains why $\tTh$ and $t_{H}$ grow exponentially with the system size, which makes the detection of the correlation hole experimentally challenging. To handle this problem, we deal with small system sizes. The other issue for both quantities is the lack of self-averaging~\cite{Schiulaz2020,Torres2020a,Torres2020b}, which requires the use of ensemble averages for any system size. To avoid being too demanding for the experiments, we use ensembles that are relatively small, although large enough for revealing the ramp.

\section{Dynamical manifestations of many-body quantum chaos} 
\label{secIII}

We start the analysis of the correlation hole with
the spin-1/2 Heisenberg model and show that dynamical manifestations of many-body quantum chaos can be detected in a chain with only 6 sites. Moving next to the long-range interacting Ising model, we verify that the correlation hole is not as clearly discernible for small system sizes as in the Heisenberg model.

\subsection{Disordered spin-1/2 Heisenberg model}
Spin-1/2 Heisenberg models and related models can be experimentally realized with NMR platforms~\cite{Wei2018,Peng2023}, inelastic neutron scattering~\cite{Scheie2022}, cold atoms~\cite{Hild2014},  Rydberg atoms~\cite{Signoles2021}, ion traps~\cite{Kranzl2023}, and quantum dots~\cite{Diepen2021}. In the presence of onsite disorder, this model has been   extensively used in studies of many-body localization~\cite{SantosEscobar2004,Santos2005loc,Nandkishore2015}.

We consider a 1D isotropic spin-1/2 Heisenberg (XXX)  model with nearest-neighbor couplings and open boundary conditions  described by the Hamiltonian 
\begin{align}
	\label{eq_MBL}
	\hat{H}_\mathrm{XXX} &= \frac{1}{2}\sum_{k = 1}^{L} h_k \hat{\sigma}_k^z + \frac{J}{4} \sum_{k = 1}^{L-1} \vec{\sigma}_k\cdot \vec{\sigma}_{k+1} ,
\end{align}
where $L$ is the chain size, $\vec{\sigma}_k\equiv \{ \hat{\sigma}^x_k, \hat{\sigma}^y_k, \hat{\sigma}^z_k \}$ are the Pauli operators on the $k$th site, the random Zeeman splittings $h_k$ are uniformly distributed within $[-W, W]$, and the coupling strength $J=1$. 
The Hamiltonian conserves the total spin in the $z$-direction, ${\cal S}_z = \sum_k \hat{\sigma}_k^z/2$, so the Hamiltonian matrix consists of $L+1$ mutually decoupled diagonal blocks. We choose $L$  to be even and  work in the largest subspace, where ${\cal S}_z =0$ and the Hilbert-space dimension is $D=L!/(L/2)!^2$. 

In the absence of disorder, $W = 0$, the XXX model is integrable and solvable via the Bethe ansatz~\cite{Bethe1931}.  When the disorder strength $W\sim J$, the system is chaotic, thus presenting correlated eigenvalues~\cite{Jacquod1997, Avishai2002,Santos2004}. For $W\gg J$, the spectra of finite systems show Poisson statistics~\cite{Santos2004} suggesting localization.

To study the dynamics, the system is initially far from equilibrium. It is prepared in an eigenstate of 
$
 \hat{H}_z =  \sum_{k=1}^L h_k \hat{\sigma}_k^z/2 + J \sum_{k=1}^{L-1} \hat{\sigma}_k^z\hat{\sigma}_{k+1}^z/4   
$,
which represents the unperturbed part of the total Hamiltonian $\hat{H}_\mathrm{XXX}$. These initial states have on each site a spin pointing up or down in the $z$-direction,  such as  the N\'eel state, $|\uparrow \downarrow \uparrow \downarrow...\rangle$, and they can be experimentally prepared. We choose initial states with energy in the middle of the spectrum, $\langle \Psi(0) |\hat{H}_\mathrm{XXX} |\Psi(0) \rangle \sim 0$, where strong hybridization of neighboring eigenstates allows chaos to develop for sufficiently small disorder strength.

\begin{figure*}[!t]
	\centering
	\includegraphics[width=0.8\textwidth]{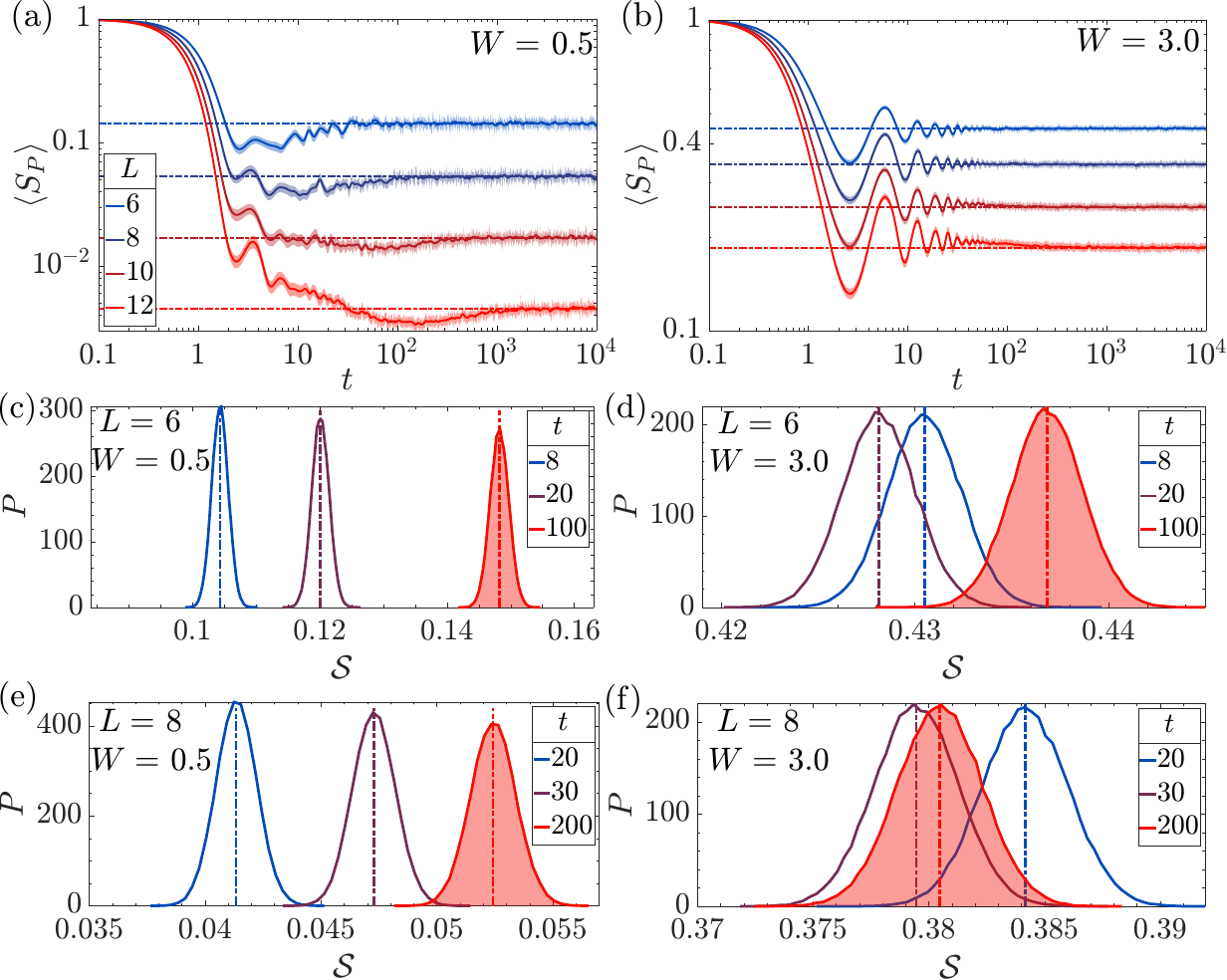}
	\caption{(a)-(b) Evolution of the survival probability under the Heisenberg spin-1/2 model in Eq.~\eqref{eq_MBL} in the (a) chaotic ($W = 0.5$) and (b) localized  ($W = 3$) regime for system sizes $L$ increasing from top to bottom, as indicated the legend in (a).
    Average over 10 initial states and 50 (500) disordered realizations for $W = 0.5$ ($W = 3$). 
    All data are further smoothed with a running time-averaged through 10 consecutive points resulting in the solid lines. The shaded area around the lines denotes the 95\% confidence interval of the ensemble averaged survival probability. 
    The horizontal dot-dashed lines mark the asymptotic values, $\overline{S_P}$. Time is in the unit of $J = 1$. (c)-(f)  Shot noise experiment for the survival probability with $M=100$ measurements for each initial state (10 samples) and disorder realization (50 samples). 
    Distributions $P$ of ${\cal S}$ for different system sizes, disorder strengths, and times as indicated in the panels.  
    The shaded distribution is obtained for a time in the saturation region. Vertical lines mark the average values  $\langle {\cal S}  \rangle$ for the chosen times.
	}
	\label{fig_MBL}
\end{figure*}
%

\subsubsection{Survival probability}

In Fig.~\ref{fig_MBL}(a) and Fig.~\ref{fig_MBL}(b), we show  the evolution of the averaged survival probability for the XXX model in Eq.~\eqref{eq_MBL} in the chaotic  ($W = 0.5$) and localized ($W = 3$) phase, respectively. 
The horizontal dot-dashed lines indicate the saturation value, $\overline{S_P}$.
 Our goal is to identify a ramp below $\overline{S_P}$ for the system in the chaotic regime.

In Fig.~\ref{fig_MBL}(a), we observe that the averaged survival probability exhibits the slope-dip-ramp-plateau structure even for  system sizes as small as $L=6$ and $L=8$. The structure is particularly visible for $L=6$, because the system size is so small that the dynamics does not develop a power-law decay. In this case, the Gaussian decay, $e^{-\Gamma^2 t^2}$, is followed by the ramp, which makes the correlation hole rather evident.
For $L=6$, the minimum value of $\langle S_P(t) \rangle$ at $\tTh$ is large enough, $\langle S_P( \tTh )\rangle_\mathrm{min}\sim \mathcal{O}(10^{-1})$,
and $\tTh$ is small enough, $\tTh \sim \mathcal{O}(10)$, to be within the grasp of current experimental setups. Even the saturation time, $t_{\mathrm{H}} \lesssim 10^2$, is at the limit of what can be experimentally reached.
Notice, however, that it is not essential to run the experiment up to $\tH$. To convince oneself that chaos has been dynamically captured, it should suffice to detect the ramp, that is, to measure values of $\langle S_P(t)\rangle$ that consistently increase as time passes, following the ramp described by the $b_2(t)$ function in Eq.~(\ref{eq_b2}). 

The purpose of Fig.~\ref{fig_MBL}(b) is to make evident that even though one can find values of $\langle S_P(t)\rangle$ below the saturation line when the system in non-chaotic, these values are not consistently below $\overline{S_P}$ and they are not on a ramp described by $b_2(t)$, so they are not caused by the presence of correlated eigenvalues. We chose to show the non-chaotic case with $W\neq 0$ instead of $W=0$ (integrable point), because we could smooth the curves using averages over disorder realizations, but the discussion is valid for both cases. 

Due to its lack of self-averaging, the survival probability has to be averaged to display the correlation hole. In Fig.~\ref{fig_MBL}(a), we average over 10 initial states and 50 disorder realizations (we verified numerically that 10 random realizations suffice to reveal the ramp). We use  10 initial states and 500 disorder realizations in Fig.~\ref{fig_MBL}(b). A much larger number of realizations is needed for the convergence of the results in the localized phase. This is because in the chaotic regime,  the relative variance of the fluctuations of $\langle S_P(t) \rangle$ at long times remains constant as $L$ increases~\cite{Schiulaz2020}, while it increases with system size in the localized phase~\cite{Torres2020a}.

\subsubsection{Shot-noise experiment}

Experimentally, the averaged survival probability is measured as follows. One chooses a particular realization of the onsite disorder, prepares the system in a specific initial state $\ket{\Psi(0)}$, and after letting it evolve unitarily for a time $t$, one performs a projective measurement. Each measurement corresponds to a ``shot'' and its outcome can be either 0 or 1 \cite{Matsoukas2023pra}. For $M$ number of shots, $M_1$ is the number of times that one gets the outcome 1, where $0\leq M_1\leq M$. According to the measurement postulate of quantum mechanics,
$\lim_{M\to\infty} \dfrac{M_1}{M} = S_P(t)$, 
where $S_P(t)$ is the survival probability, as defined in Eq.~\eqref{eq_SP_def}. In a real experiment, $M$ is finite.

We repeat the procedure above for 10 initial states and 50 random realizations to get the averaged survival probability, $\langle S_P(t)\rangle $.
To emulate the experiment, we define a random variable  
\begin{align}
    \label{eq_def_xi_shot}
    s = \begin{cases}
        1  \hspace{0.2 cm} \textrm{with probability } S_P(t)\\
        0 \hspace{0.2 cm} \textrm{with probability } 1 - S_P(t)
    \end{cases}.
\end{align}
For each initial state and  disorder realization, the sequence of $M$ random numbers $s$ gives a value $\tilde{\cal S}(t) = M_1/M$. We then define ${\cal S}(t)$ as the average of $\tilde{\cal S}(t)$ over 10 initial states and 50 disorder realizations. If $M$ is very large, then ${\cal S}(t) \to  \mean{S_P(t)}$. 

For a finite $M$, every time we repeat the procedure above, we get a value of ${\cal S}(t)$ that fluctuates around $\mean{S_P(t)}$. This fluctuation is called shot noise and is proportional to $1/\sqrt{M}$. The distribution of the values of ${\cal S}(t)$ has a width $\cal{K}/\sqrt{M}$, where $\cal{K}$  can be obtained by studying the width of the distribution of ${\cal S}(t)$ as a function of $M$. 

The number of shots that are experimentally viable depends on the experimental platform that is used. 
To identify the correlation hole, the uncertainty in ${\cal S}(t)$ must be smaller than the depth of the correlation hole, $\delta = \langle S_P(\tH) \rangle - \langle S_P(\tTh) \rangle $. To resolve ${\cal S}(\tTh)$ from ${\cal S}(\tH)$ with 99.73\% certainty, we need to have $\delta > 3\mathcal{K}/\sqrt{M}$. Based on this reasoning, we estimate that $M \sim 40$ for the survival probability evolving under the XXX model with $L = 8$ and even less for $L=6$. This means that $\mathcal{O}(10^2)$ shots per initial state and realization should be sufficient to separate the minimum of $\langle S_P(t) \rangle $ from its corresponding asymptotic value. 
Therefore, we use $M = 100$, which should be overall accessible to current experiments.

In Figs.~\ref{fig_MBL}(c)-(f), we show distributions of the values of ${\cal S}(t)$ for $W=0.5$ [Figs.~\ref{fig_MBL}(c),(e)] and $W=3$ [Figs.~\ref{fig_MBL}(d),(f)] for $L=6$ [Figs.~\ref{fig_MBL}(c),(d)] and $L=8$ [Figs.~\ref{fig_MBL}(e),(f)]. We use 10 initial states, 50 disorder realizations, and $M=100$, and repeat the procedure $10^5$ times to obtain the density of ${\cal S}(t)$.
The distributions are depicted for three different times. Their choices are based on the numerical results for the dynamics for $W=0.5$, so that the shortest time is close to the point where the ramp starts, the second one is an intermediate time on the ramp, and the largest time is already in the region of the saturation of the dynamics. The distribution for this longest time is shaded. The choices of times also have support from the numerical simulation for $W=3$ to ensure that the selected times would not give the wrong impression of a ramp in the localized region.

We see that in the chaotic regime [Figs.~\ref{fig_MBL}(c),(e)], the distributions of ${\cal S}(t)$ are well separated, so that the values of the survival probability on the ramp can be reliably identified against the asymptotic value, and the averages of the distributions grow monotonically as time increases. 
In contrast, the distributions of ${\cal S}(t)$ for the localized phase [Figs.~\ref{fig_MBL}(d),(f)] partially overlap, which means that an even larger number of shots ($M>100$) would be required to distinguish the different values in time. Furthermore, the average $\langle {\cal S}(t) \rangle$ does not grow monotonically with time. In Fig.~\ref{fig_MBL}(d), $\langle {\cal S}(t) \rangle$ for $t=20$ is smaller than for $t=8$, and in Fig.~\ref{fig_MBL}(f), the shaded distribution, which is for a time already in the saturation region, has $\langle {\cal S}(t) \rangle$ smaller than that for the chosen intermediate time.  Thus, in an real experimental scenario, the combination of knowledge from the numerical simulations and a reasonable set of points in time should enable the distinction between chaos and non-chaos in chains with only $L=6$ sites.

\begin{figure}[t!]
    \centering
    \includegraphics[width=0.48\textwidth]{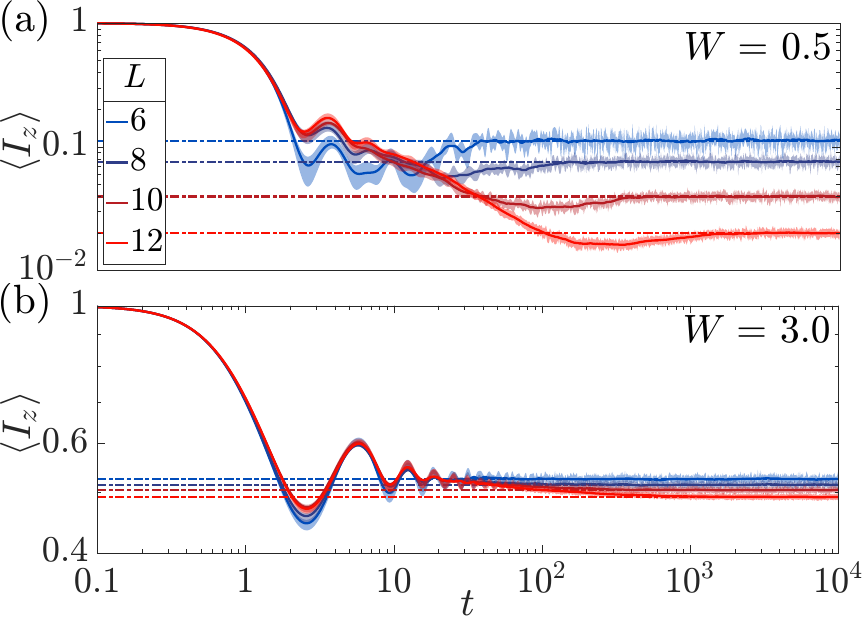}
    \caption{Evolution of the spin-autocorrelation function under the Heisenberg spin-1/2 model in Eq.~\eqref{eq_MBL} in the (a) chaotic ($W = 0.5$) and (b) localized ($W = 3$) regime for various system sizes $L$. Average and smoothing as in Fig.~\ref{fig_MBL}. Time in the unit of $J=1$. The horizontal dot-dashed lines denote the asymptotic value $\overline{I_z}$.
	}
	\label{fig:XXX_Imb}
\end{figure}
\subsubsection{Spin autocorrelation function}
In Fig.~\ref{fig:XXX_Imb}(a) and Fig.~\ref{fig:XXX_Imb}(b), we show the evolution of the averaged spin-autocorrelation function for the XXX model in Eq.~\eqref{eq_MBL} in the chaotic  ($W = 0.5$) and localized ($W = 3$) phase, respectively.
The data are averaged and smoothed as in Fig.~\ref{fig_MBL}. Similarly to the case of the survival probability, a correlation hole is visible in the chaotic regime [Fig.~\ref{fig:XXX_Imb}(a)] even for small system sizes. However, 
despite the experimental advantage of the spin autocorrelation function as a local quantity, the numerical results  are noisier than for the survival probability, because $I_z(t)$ can also have negative values. This means that, compared with the survival probability,  a larger number of measurements would be  required to experimentally reproduce the results in Fig.~\ref{fig:XXX_Imb} and to distinguish the ramp from the saturation value.

\subsection{Disordered long-range interacting Ising model}

We now analyze a disordered chain of spin-$1/2$ particles with long-range interaction, as those experimentally realized with ion traps~\cite{Jurcevic2014,Richerme2014,Smith2015}. The model has onsite disorder and open boundary conditions, being described by the Hamiltonian
\begin{align}
	\label{eq_LRT}
	\Hlr &= \frac{1}{2}\sum_{k = 1}^{L} (B+D_k) \hat{\sigma}_k^z +  \sum_{j<k} \frac{J}{(k-j)^\alpha} \hat{\sigma}_j^x \hat{\sigma}_k^x ,
\end{align}
where $B$ indicates a constant magnetic field in the transverse direction, $D_k$ is uniformly distributed within $[-W, W]$, and $\alpha$ controls the range of the spin-spin interaction. To link with the experiment in~\cite{Smith2015}, we choose $B = 2$ and $\alpha = 1.1$. As in the XXX model, we take $J = 1$ to fix the energy unit and $W = 0.5$ to access the chaotic regime. For $W = B = 0$, $\Hlr$ describes the Sherrington-Kirkpatrick spin glass model~\cite{Binder1986}, while the infinite-range interaction limit ($\alpha = 0$) yields the Lipkin-Meshkov-Glick model, which is an ideal test-bed for phenomena like excited-state quantum phase transition~\cite{Santos2016,Cejnar2021,Defenu2023} and quantum scars~\cite{Lerose2023}. 

\begin{figure}[h]
    \centering
    \includegraphics[width=0.48\textwidth]{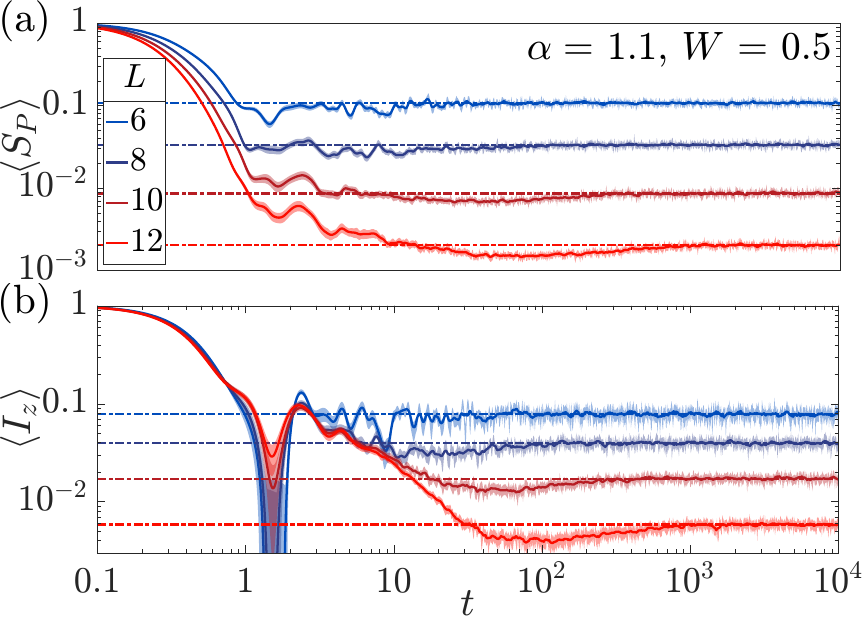}
	\caption{Evolution of the (a) survival probability and (b) spin-autocorrelation function under the long-range interacting spin-1/2 Ising model in Eq.~\eqref{eq_LRT} in the chaotic regime with $B = 2$, $W = 0.5$, and $\alpha = 1.1$ for various system sizes, $L$. The data are averaged over 10 initial states and 50 disorder realizations, and are further smoothed with a  moving time-average (solid lines), as described in Fig.~\ref{fig_MBL}. Time in the unit of $J=1$. The horizontal dot-dashed lines denote the asymptotic values.
 }
	\label{fig:LRT}
\end{figure}

The Hamiltonian $\Hlr$  decomposes into two symmetry sectors, one spanned by spin configurations with an odd number of up-spins in the $z$-direction and the other with an even number of up-spins. To maximize our access to the center of the spectrum, we use the sector with an even number of up-spins when $L/2$ is even, and the odd sector otherwise. The initial states are once again product states in the $z$-direction, which are experimentally accessible.

In Fig.~\ref{fig:LRT}, we show the evolution of the averaged survival probability [Fig.~\ref{fig:LRT}(a)] and the averaged spin-autocorrelation function [Fig.~\ref{fig:LRT}(b)] for the long-range interacting Ising model in Eq.~\eqref{eq_LRT} in the chaotic regime ($\alpha=1.1$, $W = 0.5$). In contrast with the Heisenberg model, the correlation hole for $\langle S_P(t)\rangle$ becomes clearly visible only when $L>8$ and for $\langle I_z(t)\rangle$, it needs $L\geq 8$, which makes the experimental detection of the correlation hole more challenging for the long-range interacting Ising model.

\section{Quantum circuit and many-body quantum chaos detection} 
\label{Sec:circuit}

\begin{figure*}[t]
	\centering
	\includegraphics[width=\textwidth]{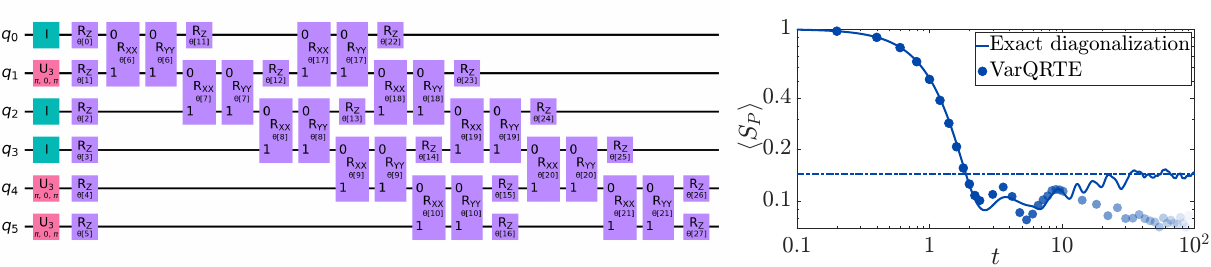}
	\caption{The left panel provides the quantum circuit used to obtain the evolution of the survival probability for the chaotic Heisenberg model with $L=6$ sites shown on the right panel with dark and light blue circles. The solid line is obtained with exact diagonalization. The horizontal dot-dashed line indicates the saturation value of the exact dynamics.
 } 
	\label{fig:circuit}
\end{figure*}

With the analysis in the previous section, one concludes that the most promising case for the experimental detection of the correlation hole is the survival probability evolving under the Heisenberg model with $L=6$ sites. In this section, we provide a quantum circuit for this case and run it in the IBM FakeSherbrooke provider~\cite{footSher}, where noise is included. Our results show that the circuit captures the beginning of the ramp.

The quantum algorithm employed is the Variational Quantum Real Time Evolution (VarQRTE) algorithm \cite{Yuan2019theoryofvariational}. Variational quantum simulation methods consist of encoding a quantum state in a parameterized circuit, and then mapping its evolution into the evolution of the parameters that control the circuit. The VarQRTE algorithm, in particular, approximately evolves an input state in a restricted subspace of the total Hilbert space, which is  defined by the ansatz of parameters used.  In the case of the Heisenberg model, the excitation preserving symmetry of the Hamiltonian motivates the use of an excitation preserving ansatz, which is naturally provided in Qiskit under qiskit.circuit.library.ExcitationPreserving \cite{footPreserve}. 

The quantum circuit that we use is shown on the left side of Fig.~\ref{fig:circuit}. Following the Qiskit notation, $q_k$ indicates the $k$th spin-1/2. In the circuit, U$_3 (\pi, 0, \pi) = \sigma_k^x$ flips the $k$th spin and I is the identity operator, so that our initial state has 3 spins pointing up in the $z$-direction and 3 spins pointing down in the $z$-direction. R$_Z (\theta) = \exp( - i \frac{\theta}{2} \sigma_k^z) $ indicates the rotation of the $k$th spin about the $z$ axis. R$_{XX} (\theta) = \exp( - i \frac{\theta}{2} \sigma_k^x \sigma_{k+1}^x)$ and R$_{YY} (\theta) = \exp( - i \frac{\theta}{2} \sigma_k^y \sigma_{k+1}^y)$ involve only neighboring spins and both gates have the same parameter to keep the evolution in the ${\cal S}_z=0$ subspace.

Using VarQRTE,  the circuit depth of the quantum circuit necessary to simulate the evolution can be chosen by the user to ensure that it is within hardware and noise constraints.  Therefore, the quantum circuits can be made shallower, at the cost of decreasing accuracy.  Additionally, simulation of long times does not require deep quantum circuits as Trotterization requires. Instead, the gate depth of the ansatz is fixed at the start of the algorithm.  For these reasons, VarQRTE is a promising noisy intermediate-scale quantum (NISQ) algorithm for detecting many-body quantum chaos.

The ansatz that we use has two layers (the two diagonal lines of gates in the circuit) and linear entanglement in order to minimize gate depth. The circuit is run in the IBM FakeSherbrooke provider, which mimics the real corresponding noisy IBM quantum system. Despite the limited gate depth of the employed ansatz and the presence of noise, the simulation is able to detect the beginning of the correlation hole, as seen on the right panel of Fig.~\ref{fig:circuit}. This figure shows with blue  circles the result produced by the circuit for the evolution of the survival probability under the chaotic Heisenberg model, and the solid line represents the result obtained with exact diagonalization. There is reasonable agreement between the two curves up to $t \sim 12$ (dark blue circles), which allows for capturing the beginning of the ramp. Beyond this point (light blue circles), noise kicks in and the curves diverge. This result is very encouraging. Ways to further improve the circuit will be discussed in a forthcoming paper.

\section{Single-excitation case} 
\label{secIV}

\begin{figure*}[t!]
	\centering
	\includegraphics[width=0.85\textwidth]{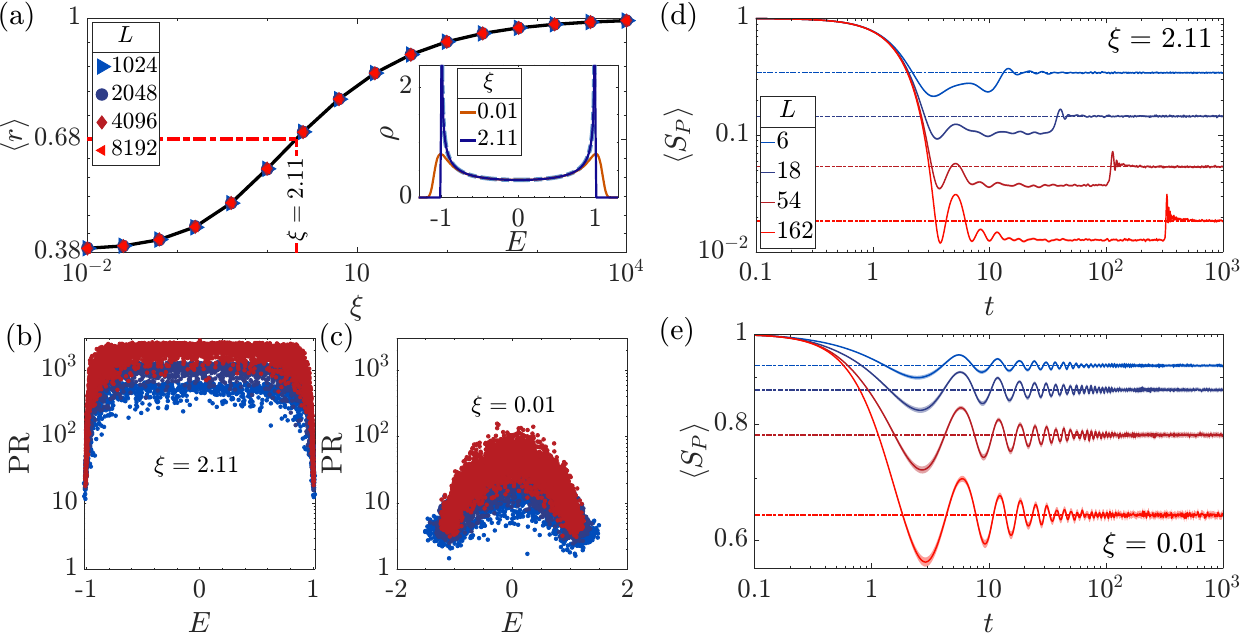}
	\caption{Results for the spin-1/2 model in Eq.~(\ref{eq_Anderson}) with a single excitation. (a) Average ratio of level spacing, $\langle r \rangle$, as a function of $\xi \propto (W^2 L)^{-1}$ for different system sizes, $L$. The red dashed line indicates the value $\xi=2.11$  where $\mean{r}\approx 0.68$, which is the average ratio observed in GSE. The inset shows the density of states $\rho(E)$ averaged over $10^4$ disorder realizations for two different values of $\xi$ and $L = 8192$; the dashed line indicates the analytical expression of the density of states for $\xi\gg 1$.
    (b)-(c) Participation ratio as a function of energy from a single disorder realization for (b) $\xi = 2.11$ and (c) $\xi = 0.01$.
    (d)-(e) Survival probability for (d) $\xi = 2.11$ and (e) $\xi = 0.01$ for different system sizes. The  initial state has the excitation on the first site of the chain. The data are averaged over 200 disordered realizations and further smoothed with a moving time average. Time is in the unit of $J=1$. The horizontal dot-dashed lines denote the asymptotic values of $\langle S_P (t) \rangle$. 
	}
	\label{fig:A1D}
\end{figure*}

To provide one more case in which  dynamical manifestations of spectral correlations could be experimentally detected with available capabilities, we resort to a system with a single excitation described by the following Hamiltonian 
\begin{align}
	\label{eq_Anderson}
	\begin{split}
		\hat{H} &= \frac{1}{2} \sum_{k=1}^L h_k \del{\hat{\sigma}^z_k + \mathbb{I}}  + \frac{J}{4}\sum_{k=1}^{L-1} (\hat{\sigma}^x_k \hat{\sigma}^x_{k+1} + \hat{\sigma}^y_k \hat{\sigma}^y_{k+1}),
	\end{split}
\end{align}
where $h_k$'s are random Zeeman splittings uniformly distributed in $[-W,W]$. This Hamiltonian is similar to that in Eq.~\eqref{eq_MBL}, where only nearest-neighboring couplings are present, but it does not have the Ising interaction. In the thermodynamic limit, the system in Eq.~\eqref{eq_Anderson} exhibits Anderson localization for any finite disorder strength~\cite{Anderson1958} with a localization length for eigenstates  away from the edges of the spectrum given by $l\approx 6.5653/W^2$ \cite{Izrailev1998}. To investigate finite system sizes, one can then take the scaled localization length $\xi = 6.5653/(W^2L)$ as a parameter.

The system in Eq.~\eqref{eq_Anderson} is not chaotic, but if the localization length is larger than the system size, the energy levels are correlated and show Wigner-Dyson distribution~\cite{Sorathia2012,Torres2019b,Elkamshishy2021, Das2023a, Das2024, Das2025ArxivPower, Das2019, Das2022c}. Despite being just a finite-size effect, these spectral correlations also get dynamically manifested~\cite{Torres2019b} and could then be experimentally detected.

Figure~\ref{fig:A1D}(a) provides the results for the analysis of short-range correlations done with the ratio~\cite{Oganesyan2007, Atas2012},
\begin{equation}
r_n = \min \left( \tilde{r}_n, \frac{1}{\tilde{r}_n}\right), \quad \text{where} \quad \tilde{r}_n=\frac{E_{n+1} - E_n}{E_{n} - E_{n-1}}.
\end{equation}
The figure shows the average $\langle r\rangle$ over $r_n $ for states in the middle of the spectrum as a function of $\xi$. As $\xi$ increases and the localization length becomes larger than the system size, the level spacing distribution moves from  Poisson (absence of correlations) to distributions that indicate stronger and stronger level repulsion, 
reaching statistics equivalent to those for Gaussian orthogonal ensemble (GOE), Gaussian unitary ensemble (GUE), Gaussian sympletic ensemble (GSE), and  beyond (picket-fence). 

When $\xi=2.11$ [dashed line in Fig.~\ref{fig:A1D}(a)], the average ratio $\langle r \rangle$ is approximately that observed in the GSE. This is the value that we use for the analysis of the survival probability, because among the ensembles, the GSE leads to the lowest values of $\langle S_P (t) \rangle$ below the saturation line of the dynamics. We compare the results with those for $\xi=0.11$, for which the level spacing distribution is Poissonian.

The comparison of the participation ratio,
\begin{equation}
    \text{PR}_n = \frac{1}{\sum_k |\langle k|E_n \rangle|^4},
\end{equation}
as a function of $E_n$
 for $\xi=2.11$ [Fig.~\ref{fig:A1D}(b)] and $\xi=0.01$ [Fig.~\ref{fig:A1D}(c)] further corroborates that the finite system in Eq.~\eqref{eq_Anderson} is delocalized when $\xi$ is large. Notice that the density of states, shown in the inset of Fig.~\ref{fig:A1D}(a), is not Gaussian as in many-body systems, although the most delocalized states are still those in the middle of the spectrum.

To investigate the evolution of the survival probability, we prepare the system in a initial state where the excitation is on the first site of the chain. This is done, because for $W\rightarrow 0$, the shape of the LDOS for this state is semicircular~\cite{Torres2019b}, which brings us closer to the case of
Fig.~\ref{fig:GOE} and should facilitate the visibility of the dip below $\overline{S_P}$.

We show the evolution of the averaged survival probability obtained for $\xi=2.11$ in  Fig.~\ref{fig:A1D}(d) and for $\xi=0.01$ in Fig.~\ref{fig:A1D}(e). The behavior in the two panels is completely different.  The survival probability in Fig.~\ref{fig:A1D}(d) exhibits a ``correlation plateau'' at $\langle S_P (t) \rangle \sim 2/L$, which is below the saturation plateau at $\overline{S_P } \sim 3/L$ \footnote{The spectrum of this model for $\xi=2.11$ presents a combination of GSE-like statistics, in the case of short-range correlations, and GOE-like statistics for long-range correlations. It has been shown analytically with full random matrices that the minimum value of $\langle S_P (t) \rangle = 2/(\beta D)$ and the saturation value $\overline{S_P}=(\beta+2)/(\beta D+2)$, where $\beta =1, 2, 4$ for GOE, GUE, and GSE~\cite{Alhassid1992}.}, while $\langle S_P (t)\rangle $ in Fig.~\ref{fig:A1D}(e) simply oscillates around $\overline{S_P }$.

The results in Fig.~\ref{fig:A1D}(d) indicate that even though the long-time dynamics of the survival probability for the Hamiltonian in Eq.~(\ref{eq_Anderson}) is not described by the $b_2(t)$ function in Eq.~(\ref{eq_b2}), as in truly chaotic systems, one can still capture dynamical manifestations of spectral correlations in spin systems with a single excitation. In this case, since the saturation (Heisenberg) time  scales linearly with the chain size ($\tH \propto L/\Gamma$, where $\Gamma \sim 1/2$), it should be viable to experimentally run the entire evolution up to saturation. This simple scenario could serve as a first step towards the  experimental detection of many-body chaos.

\section{Conclusion} \label{secV}

Experimental advances in engineering pure initial states, preserving quantum coherences for long times, and monitoring the time evolution of quantities of interest to many-body quantum systems set the stage for the direct observation of dynamical manifestations of many-body quantum chaos, specifically of the onset of the slope-dip-ramp-plateau structure (correlation hole), that is typical of systems with correlated eigenvalues. Several existing quantum simulators and quantum computers  can be generally described by interacting spin-1/2 models. This motivated our analysis of the emergence of the correlation hole in the survival probability and in the spin autocorrelation function evolved under two chaotic spin-1/2 models: the 1D disordered Heisenberg model with nearest-neighbor couplings and the 1D  disordered long-range interacting Ising model in a transverse field.

We concluded that the averaged survival probability evolved under the disordered Heisenberg chain with only $6$ sites offers the best prospect for the detection of many-body quantum chaos with available experimental resources. Furthermore, our analysis of the shot-noise experiment for the survival probability indicated that measurements at some few times within the correlation hole should suffice for inferring its presence.

Inspired by the promising results obtained with exact diagonalization, we employed the VarQRTE algorithm to simulate with a quantum circuit the time evolution of the survival probability for the 6-site Heisenberg model. We used a fake IBM provider to run the circuit and, despite the presence of noise, we were able to capture the beginning of the correlation hole. This result should motivate further improvements to the circuit to make it shallower and to mitigate errors, which we intend to pursue in a forthcoming work.

The detection of dynamical manifestations of spectral correlations, but not necessarily many-body quantum chaos, could also be achieved with quantum systems of few excitations or few degrees of freedom. We presented the case of a spin-1/2 model with a single excitation, where correlated eigenvalues are due to finite-size effects. These correlations get manifested in the dynamics at times that are shorter and at values of the survival probability that are larger than what we have for many-body systems of the same length. Another model that could be used for the experimental detection of quantum chaos is the Dicke model, which describes a set of $N$ spin-$1/2$ particles collectively interacting with a single-mode field and also exhibits a correlation hole~\cite{Lerma2019,Villaseor2020}. Because the interaction is collective, the system has only two degrees of freedom.

We close this work with a brief discussion about the case of noisy systems.
Recent investigations of the spectral form factor in energy dephasing scenarios~\cite{VallejoFabila2024, Xu2021,Cao2022,Matsoukas2023pra,Matsoukas2023quantum}, non-Hermitian Hamiltonians~\cite{Cornelius2022,Matsoukas2023}, and parametric quantum channel models~\cite{Matsoukas2023quantum} suggest that weak interactions with the environment reduce the need for ensemble averaging. This implies that some noise can be beneficial for the experimental detection of the correlation hole, a subject that deserves further investigation.

\begin{acknowledgments} 
	A.~K.~D. is supported by the Fulbright-Nehru Grant No.~2879/FNDR/2023-2024. D.A.Z.-H. and E.J.T.-H. are supported by CONAHCYT through project No. CF-2023-I-1748. D.C. and V.S.B. are funded by the NSF CCI grant
(Award No. 2124511). A.d.C. is supported by 
the Luxembourg National Research Fund (FNR), grant reference [17132054].  E.J.T.-H. is  grateful for financial support from VIEP-BUAP, project No. 00165-2023. D.A.Z.-H. and E.J.T.-H. are grateful to LNS-BUAP for their supercomputing facility. L.F.S. is supported by the NSF Grant No. DMR-1936006. This research was supported in part by grants NSF PHY-1748958 and PHY-2309135 to the Kavli Institute for Theoretical Physics (KITP). For the purpose of open access, the authors have applied a Creative Commons Attribution 4.0 International (CC BY 4.0) license to any Author Accepted Manuscript version arising from this submission.
\end{acknowledgments}

\bibliography{WN10103} 

\end{document}